\documentclass[prl,aps,twocolumn,amssymb,floatfix,superscriptaddress,notitlepage]{revtex4-2}
 \usepackage{times}
\usepackage{graphicx}
\usepackage{amsmath}
 \usepackage{amstext}
\usepackage{amssymb}
\usepackage{fancybox}
\usepackage{xfrac}
\usepackage[colorlinks,citecolor=blue]{hyperref}
\usepackage{graphicx}
\usepackage{amsmath}
\usepackage{amstext}
\usepackage{amssymb}
\usepackage{amsfonts}
\usepackage{longtable,booktabs}
\usepackage{hyperref}\usepackage{url}
\usepackage{subfigure}%
\usepackage{dsfont}
\usepackage{booktabs}
\usepackage{amsbsy}
\usepackage{dcolumn}
\usepackage{amsthm}
\usepackage{bm}
\usepackage{esint}
\usepackage{multirow}
\usepackage{hyperref}
\hypersetup{
    colorlinks=true,
    linkcolor=magenta,
    filecolor=magenta,
    urlcolor=magenta,
}
\usepackage{cleveref}

\usepackage{mathrsfs}
\usepackage{amsfonts}
\usepackage{amsbsy}
\usepackage{dcolumn}
\usepackage{bm}
\usepackage{multirow}
\usepackage{color}

\newcommand{\comments}[1]{}
\def\Z{\mathbb{Z}}

  \usepackage{extarrows}
\usepackage{datetime}

\usepackage{comment}
\usepackage[super]{nth}

 \usepackage{varwidth}

\usepackage[english]{babel}
\usepackage[utf8x]{inputenc}
\usepackage[T1]{fontenc}
\usepackage{epstopdf}
\usepackage{float}
 \usepackage{amssymb}
\usepackage[force]{feynmp-auto}
\usepackage[T1]{fontenc} 
\usepackage{amsmath}
\usepackage{graphicx}
\usepackage[colorinlistoftodos]{todonotes}
\usepackage{ulem}
\usepackage{float}

\newcommand{\bpm}{\begin{pmatrix}}
\newcommand{\epm}{\end{pmatrix}}

\def\Z{\mathbb{Z}}

\begin{document}

\title{Parity violation as enforced symmetry breaking in 3D fermionic topological order}
  
\author{Shang-Qiang Ning}
\affiliation{Department of Physics, The Chinese University of Hong Kong, Shatin, New Territories, Hong Kong, China} 
\affiliation{Department of Physics, The Hong Kong University of Science and Technology, Clear Water Bay, Kowloon, Hong Kong, China} 
\author{Yang Qi}
\email{qiyang@fudan.edu.cn}
\affiliation{Center for Field Theory and Particle Physics, Department of Physics, Fudan University, Shanghai 200433, China}
\affiliation{State Key Laboratory of Surface Physics, Fudan University, Shanghai 200433, China}

\author{Chenjie Wang}
\email{cjwang@hku.hk}
\affiliation{Department of Physics and HK Institute of Quantum Science \& Technology, The University of Hong Kong, Pokfulam Road, Hong Kong, China}

\author{Zheng-Cheng Gu}
\email{zcgu@phy.cuhk.edu.hk}
\affiliation{Department of Physics, The Chinese University of Hong Kong, Shatin, New Territories, Hong Kong, China}

\begin{abstract}
Symmetry can be intrinsically broken in topological phases due to inherent incompatibilities, a phenomenon known as enforced symmetry breaking (ESB) in the framework of topological order. In our previous work, we developed a systematic framework to understand ESB within 2D invertible topological order. Meanwhile, the origin of parity violation in the Standard Model remains one of the most profound mysteries in physics, with no clear explanation to date. In this study, we explore the ESB of parity symmetry by three-dimensional fermionic topological order (fTO), offering potential insights into the origins of parity violation.
As the simplest example, here we consider an fTO related to the intrinsic interacting fermionic SPT phase protected by $\Z_2^f\times \Z_2\times \Z_8$ symmetry in three dimensions. We show that time-reversal symmetry (TRS) with $\mathcal{T}^2=1$ on  physical fermions is incompatible with such fTO; then, through the so-called crystalline equivalence principle, we show that the  parity symmetry is also incompatible with it. In comparison, conventional TRS with $\mathcal{T}^2=\mathcal{P}_f$ remains compatible to this fTO. We also discuss a general framework to study the ESB phenomenon for 3D fTO.
\end{abstract}

\maketitle

\textit{Introduction} --- Symmetry and its breaking are at the very heart of modern physics—from phenomena such as Bose-Einstein condensation, magnetization, and superconductivity in condensed matter to the Higgs mechanism in high-energy physics \cite{Landau1937obd}. Yet, the discovery of the fractional quantum Hall effect ushered in an era of topological phases that transcend the traditional Landau paradigm \cite{Klitzing1980,Tsui1982,Laughlin1983}. The intricate interplay between symmetry and topology has, in turn, given rise to a myriad of new concepts, notably symmetry-protected topological (SPT) order and symmetry-enriched topological (SET) order, which have attracted extensive recent attention. . 

However, symmetry does not always align compatibly with a topological order (TO). In certain situations, a symmetry can be  incompatible with a given TO, which means there does not exist a Hamiltonian whose ground state realizing the given TO is invariant under that symmetry. Such an incompatibility is called enforced symmetry breaking (ESB).  It is well known that the 1D Kitaev chain (KC)—the sole example of intrinsic topological order in one dimension—is incompatible with a time-reversal symmetry (TRS) defined by $\mathcal{T}^2=\mathcal{P}_f$ while compatible with $\mathcal{T}^2=1$\cite{Fidkowski11}. 
The ESB can be determined by the inherent  topological nature of the TO,  without necessitating the explicit ground state wavefunctions.
So it is a kind of intrinsic symmetry breaking rooted in the universal many-body entanglement structure of the TO. 
While the renowned LSM theorem can limit possible ground states \cite{LSM61, Affleck86, Affleck88, Masaki97, Masaki00, Hastings04, Meng16} by only forbidding the gapped symmetric featureless state,
the ESB, however, imposes stricter  constraints on ground states than LSM-type theorems by ruling out certain TOs. 

In the previous work Ref.\cite{2DESB}, we systematically studied the ESB  for 2D invertible fTO.  Herein, we venture into the ESB phenomenon of 3D fTO. As there is no invertible fTO in 3D, we have to consider intrinsic fTO.  
In this work, we will focus on  3D ESB of TRS.
We show that TRS is intrinsically broken by the presence of certain 3D fermionic TO (fTO) characterized by exotic non-Abelian three-loop braiding statistics.
Such an fTO is closely related to an intrinsic interacting fermionic SPT (fSPT) order, which is specifically protected by symmetry $K_f=\Z_2^f\times \Z_2\times \Z_8$, characterized by non-Abelian three-loop braiding statistics \cite{Zhou21}. This fTO is obtained by gauging $\Z_2\times\Z_8$ in the fSPT state, and will be referred to as Ising fTO. 
As we shall show, this $K_f$ fSPT is incompatible with $\mathcal{T}^2=1$ while compatible with $\mathcal{T}^2=\mathcal{P}_f$.  
Its incompatibility with the $\mathcal{T}^2=1$ TRS becomes exactly the ESB of TRS by this fTO.  To our knowledge, this is the first and simplest example of TRS incompatibility of 3D fTO. 
This example readily extends to the realm of parity symmetry through the crystalline equivalence principle, while the concept of ESB unveils a novel pathway for comprehending parity violation within the Standard Model (SM), under the assumption that its vacuum acquires emergent generalized symmetry\cite{GS1,GS2,GS3} characterized as certain 3D Ising fTO.
We will also discuss the general framework for understanding 3D ESB phenomena.

\begin{figure}[t]
       \centering
   \includegraphics[width=8cm]{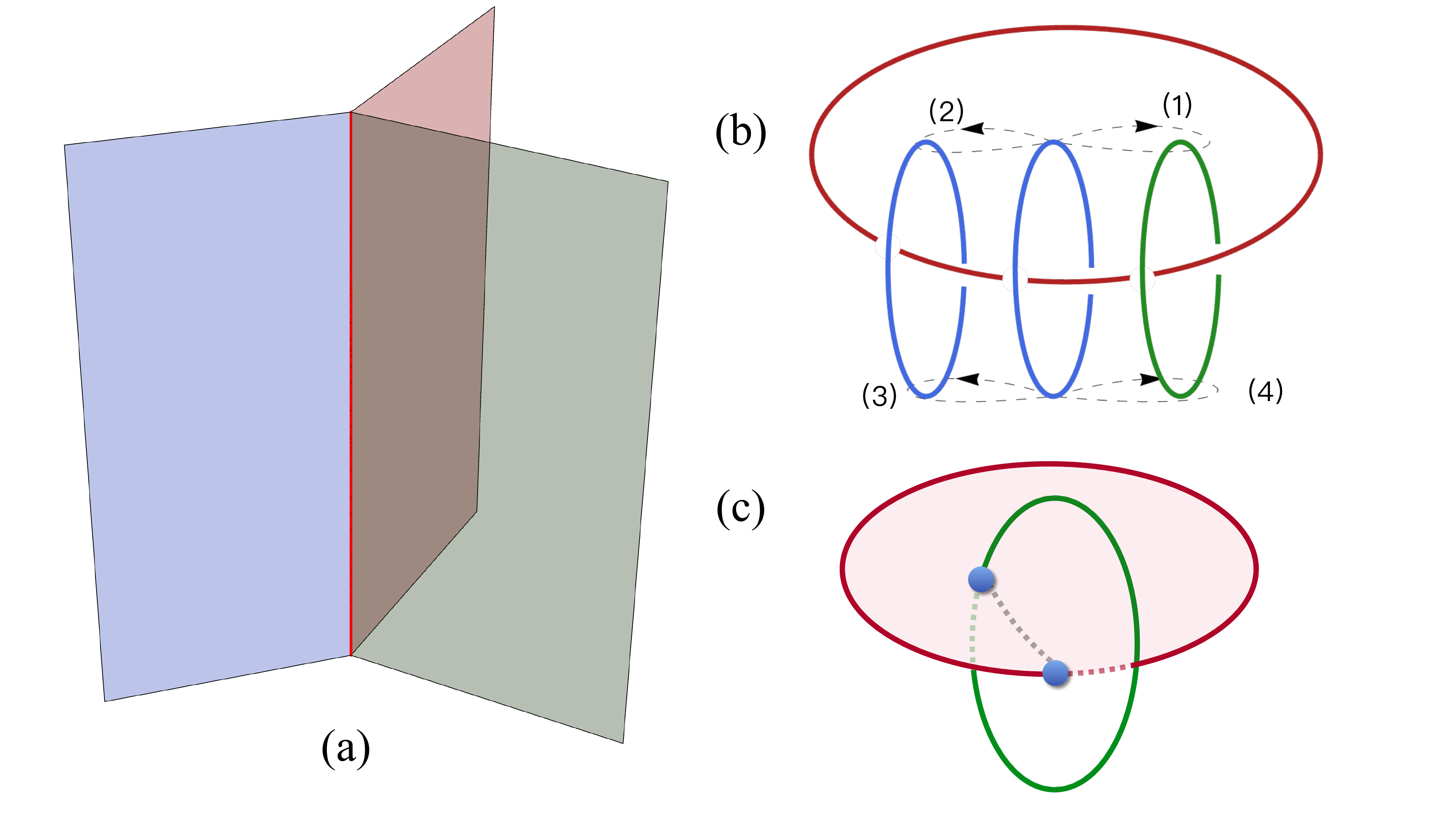} 
      \caption{ The physics of 3D fTO and corresponding fSPT. (a) The fSPT is characterized by KC decoration on the intersection line of $\Z_2$ and $\Z_8$ domain wall.  (b) is about non-Abelian three loop braiding statistics. Blue loops represent two $\Omega_0$ loop excitations, green one represents the bosonic $\Z_2$ loop and the red base loop is the $\Z_8$ loop. We  braid the middle $\Omega_0$ loop along the path labeled by (1), (2), (3) and (4) consecutively.  Note that  any of the four loops in (b) can be base loop and the other three exhibit the same non-Abelian three loop braiding statistics. (c) For two linked $\Omega_1$ and $\Omega_2$,  each carries a MZM at the point of the loop intersecting with the other supporting membrane, denoted as blue dot.  }
   \label{fig:NATL}
\end{figure}

\textit{3D fSPT and fTO} --- Let us briefly review the $K_f$ 3D fSPT and the corresponding gauged fTO.
An elegant way to understand the SPT is using the decoration domain wall construction\cite{Xie14DDW,QR18,QR21}.
In the 3D bulk with $\Z_2\times \Z_8$ symmetry, there are 2D $\Z_8$ and $\Z_2$ domain walls (DW), denoted $\mathbf{g}$ and $\mathbf{h}$, respectively.  This fSPT is constructed by decorating the KC on the intersection line of $\mathbf{g}$ and $\mathbf{h}$ DWs, whose  coherent fluctuation gives rise to the fSPT wavefunction (see Fig.\ref{fig:NATL}a). The decoration is characterized by a function that depends on two group elements in $\Z_2\times \Z_8$, called the two-cocycle $n_2 $ that can take a value of 0 or 1, indicating the absence or presence of KC \cite{QR18}. On the other hand, it can  be equivalently viewed as the coherently decorated 2D root 
$\Z_2^f \times \Z_2$ fSPT on 2D $\mathbf{g}$ DW  \footnote{The intersection of the $\mathbf{g}$- and $\mathbf{h}$-DW can be viewed as  a 1D  $\mathbf{h}$-DW on the 2D  $\mathbf{g}$-DW, where the proliferation of the (1D)  $\mathbf{h}$-DW decorated by the KC leads to the 2D root  $\Z_2^f\times \Z_2$  fSPT. }. The classification of 2D  $\Z_2^f\times \Z_2$   fSPT  is $\Z_8$ \cite{Z2fSPT1,Z2fSPT2,Z2fSPT3,Z2fSPT4}, which can be denoted as $[n]$ with $n=0,1,..7$, realizing by the stacking of $n$ copies of the root phase $[1]$.

Upon gauging $\Z_2\times \Z_8$  in this fSPT (promoting the global $\Z_2\times \Z_8$ symmetry to be a local gauge symmetry), we obtain a 3D fTO, dubbed Ising fTO,  with two elementary particle-like topological  excitations: gauge charges $\mathsf{e}_{1,2}$, together with two elementary loop excitations $\Omega_{1,2} $, containing  unit flux of  $\Z_2$ and $\Z_8$ gauge group, respectively. Besides, there is also  fermionic excitation that can be created by local operators and is usually called the transparent fermion $f$. General loop excitations contain the gauge flux and gauge charge. The flux of loop excitations forms the fusion group $ \Z_2\times \Z_8$. 
This fTO has many exotic properties. First, when $\Omega_1$ is linked to the base loop $\Omega_2$, its topological spin $\theta_{\Omega_1;\Omega_2}$ satisfies $\Theta_{\Omega_1;~\Omega_2}=2\theta_{\Omega_1;~\Omega_2}=\frac{\pi}{4}$ mod $\frac{\pi}{2}$.  Secondly, for two linked $\Omega_1$ and $\Omega_2$,  each carries a Majorana zero mode (MZM), indicating the non-Abelian nature (see Fig.\ref{fig:NATL}c)  \cite{QR18,Zhou21}.   Thirdly, when further gauging the fermion parity, this resulting TO has fascinating statistics among loops, one of which is called the non-Abelian Ising three-loop braiding statistics as seen in Fig.\ref{fig:NATL}b .  We note that in general the topological invariant $\Theta_{\Omega_1;~\Omega_2}=\frac{x\pi}{4}$ mod $2\pi$, and it is non-Abelian  when $x$ takes odd integer.  In addition, there is another topological invariant $\Theta_{\Omega_2;~\Omega_1}$ which can take $0$ or $\pi$  characterizing an Abelian bosonic TO of $\Z_2\times \Z_8$ Dijkgraaf-Witten models.  (For completeness, a concise overview of fSPT and associated three-loop braiding is provided in the supplementary materials).  We also note that the ESB by TRS or parity symmetry we discuss below applies to these fTOs for all odd $x$. 

\textit{TRS and its ESB in $3D$ fTO} --- Let us assume that there exists an anomaly-free TRS Ising fTO, such that TRS action can be consistently defined on topological excitations.
First, TRS may permute topological excitations, including both particles and loops.  
For loops, TRS may permute both gauge flux and gauge charge. Let us consider how it permutes the gauge flux. The permutations must preserve the gauge flux fusion structure, which implies that TRS must act as an automorphism of $\Z_2\times \Z_8$. We denote the permutation as $\rho$, with $\rho\in \text{Aut}(\Z_2\times \Z_8)\cong \Z_2\times D_4$. 

As the fermionic parity does not induce flux permutation, the automorphism must satisfy $(\rho)^2=1$, regardless of $\mathcal{T}^2=1$ or $\mathcal{P}_f$. Among all automorphisms in $\text{Aut}(\Z_2\times \Z_8)$, 12 of them satisfy this condition. They can be labeled by a triplet $\mathbf{r}=(p,q,t)$, where $p=0,1,2,3$ and $q,t=0,1$ with $qt=0$. More explicitly, under TRS action, $\rho(\Omega_1)=(\Omega_2)^{4t}\Omega_1$ and $\rho(\Omega_2)=(\Omega_2)^{2p+1}(\Omega_1)^q$ (Depending on the context, we use $\rho(\Omega)$ to denote either the permutations of the loop itself or its gauge flux).

Under TRS action, topological observables, including Aharonov–Bohm statistics and three-loop braiding statistics, are required to be preserved after a complex conjugation. With Aharonov-Bohm statistics, this requirement completely fixes the permutation of particles for a given permutation $\rho$ of gauge fluxes.

For three-loop statistics, this requirement leads to the conditions $\Theta_{\Omega_1;\Omega_2}=-\Theta_{\rho(\Omega_1);~\rho(\Omega_2)}$ and $\Theta_{\Omega_2;\Omega_1}=-\Theta_{\rho(\Omega_2);~\rho(\Omega_1)}$, where
\begin{align}
    \Theta_{\rho(\Omega_1);~\rho(\Omega_2)} &= (2p+1)\Theta_{\Omega_1;~\Omega_2}+t\Theta_{\Omega_1,\Omega_2;~\Omega_2} \nonumber\\
    \Theta_{\rho(\Omega_2);~\rho(\Omega_1)} &=\Theta_{\Omega_2;~\Omega_1}+q\Theta_{\Omega_2,\Omega_1;~\Omega_1}  \label{eq:Theta}
\end{align}  
(see End Matter for derivations). Solving the equations, we find (i) when $t=0$, $p=3$; (ii) when $t=1$, $p=3$ if $\Theta_{\Omega_1,\Omega_2;~\Omega_2}=0$ and $p=1$ if $\Theta_{\Omega_1,\Omega_2;~\Omega_2}=\pi$; 
(iii) $q=0$. Therefore, we are left with only three possibly consistent $\rho$, labeled by $\mathbf{r}=(3,0,0)$,  $(3,0,1)$, and $(1,0,1)$, respectively.

We still need to consider the local Kramers degeneracy (singlet or doublet) on those topological excitations that are invariant under TRS. Below, we show that no consistent Kramers degeneracy assignment exists for the above three permutations $\rho$ if the physical fermion $f$ carries $\mathcal{T}^2=1$. To do that, we first gauge the fermion parity so that the fTO becomes a bosonic TO where $f$ now becomes a true topological  excitation and there is one more elementary loop excitation from fermion parity in addition to $\Omega_1$ and $\Omega_2$.  The TRS with $\mathcal{T}^2=1$ or $\mathcal{P}_f$ now becomes whether $f$ carries a local Kramers singlet or doublet, denoted as $\mathcal{T}_f^2=1$ or $-1$, respectively. The permutation $\rho$ acts on gauge fluxes in the same way as above, since TRS does not permute the fermion-parity flux.

We use dimensional reduction to argue that $\mathcal{T}^2_f=-1$ for any consistent $\rho$. Let us take $\mathbf{r}=(3,0,0)$ as an example. In this case, the gauge flux of $\Omega_1$ is invariant under TRS, so there exists a TRS-invariant membrane operator that creates $\Omega_1$. Taking periodic boundary conditions in $z$ direction, inserting this membrane operator in $xy$ plane, and performing a dimensional reduction, we obtain a TRS-invariant 2D bosonic TO.  Next, we consider the fusion product of $\Omega_2$ (now particle-like in 2D) and $\rho(\Omega_2)$. Importantly, both $\Omega_1$ and $\rho(\Omega_2)$ are Majorana-type, such that $\Omega_2\times\rho(\Omega_2) = e + ef + \dots$  where the left side is TRS invariant as $\rho^2=1$ and each fusion outcome of the right side is TRS invariant. 
According to Refs.\cite{Levin_Stern12,StationQ}, the Kramers degeneracy of anyons appearing in the fusion outcome is locked to their topological spin, i.e., $\mathcal{T}^2_e =\theta_e$ and $\mathcal{T}_{ef}^2=\theta_{ef}$. Since $e$ is a boson and $ef$ is a fermion, we obtain $\mathcal{T}_f^2=\mathcal{T}_{ef}^2/\mathcal{T}_e^2=\theta_{ef}/\theta_e =-1$. Similarly, for $\mathbf{r}=(3,0,1)$ and $(1,0,1)$, we can use the TRS-invariant gauge flux of $\Omega_1(\Omega_2)^2$ to argue for $\mathcal{T}_f^2=-1$.  Thus,  we conclude that TRS  with $\mathcal{T}^2_f=1$ is incompatible with the 3D Ising fTO.

\textit{Absence of fSPT with $\mathcal{T}$-extended symmetry} --- To reconfirm the above conclusion, we provide an alternative argument.  It is widely believed that {all 3D anomaly-free SETs can be realized by partially gauging 3D  SPT phases}\cite{LanAB,LanEF}.  So, the existence of a 3D anomaly-free SET implies the existence of a corresponding 3D SPT phase. 
Accordingly, we will explore all possible $G_f$ which (i) is an extension of TRS by $K_f$ and (ii) contains both $\Z_2^f$ and $\Z_2\times \Z_8$ as normal subgroups. Note that $K_f$ being normal in $G_f$ does not necessarily imply that $\Z_2\times \Z_8$ is normal. Imposing the normality of $\Z_2\times \Z_8$ in $G_f$ allows us to gauge $\Z_2\times\Z_8$ and obtain the aforementioned TRS-symmetric fTO. We will classify all 3D fSPTs of all these $G_f$'s using a computer program \cite{Ouyang_2021}. Among all 3D fSPTs of all $G_f$'s, if none becomes the 3D Ising fTO after gauging $\Z_2\times \Z_8$, we then conclude that TRS is enforced to break by the Ising fTO. 

The structure of $G_f$ is described by: (1)  how TRS acts on the subgroup $\Z_2\times \Z_8$: $\rho_\mathcal{T}(\mathbf{g})=\mathbf{g}^{1+2p}\mathbf{h}^q $ and $\rho_\mathcal{T}(\mathbf{h})= \mathbf{g}^{4t}  \mathbf{h}$, 
where $\mathbf{r}=(p,q, t)$ is the same as above \cite{autogroup}; and (2) what $\mathcal{T}^2$ is: $\mathcal{T}^2=\nu_2^a\nu_2^b\in K_f$,  where $\nu_2^a=1$ or $\mathcal{P}_f$, and $\nu_2^b=1, \mathbf{g}^2, \mathbf{g}^4$ for $\mathbf{r}=(3,0,0)$ or $(p,1,0)$ and $\nu_2^b=1$  for other choices of $\mathbf{r}$ (after considering some physical constraints for simplification \footnote{There are in total 28 different $G_b$, which will simplify to 17 cases after these physical constraints.  However, we calculate all the 28 cases and found no case can reduce to the $K_f$ fSPT. }; see End Matter). All 12 choices of $\mathbf{r}$ are considered, not limited to the three choices compatible with three-loop statistics. 
The Kramer-singlet and -doublet fermion systems correspond to $\nu_2^a=1$ and $P_f$, respectively. 
Below we demonstrate that ESB occurs for Kramer-singlet fermions.  In this case,  $G_f=\Z_2^f\times G_b$, where $G_b$ is the group extension of $\Z_2^\mathcal{T}$ by $\Z_2\times \Z_8$ with $\rho_{\mathcal{T}}$ and $\nu_2^b$. In total, there are twelve groups with $\nu_2^b=1$ (Table \ref{tab:singlet1}) and five groups with $\nu_2^b=\mathbf{g}^{2m}$ (Table \ref{tab:Q8}).

\begin{table}[t]
\centering
\begin{tabular}{| c |c|c|c |c |c |}
\hline
\hline
 \,\, $G_f$ \,\,  & \,\, $(p,q,t)$\,\, &\,\, $\mathcal{T}^2$\,\,& Kitaev & complex fermion &\textit{  } bosonic \textit{  }  \\ 
 \hline
 $\mathsf{G}_{1}$    &(0,0,0) &1& $\times$  &  $\Z_2$ & $(\Z_2)^5$\\ 
 \hline
 $\mathsf{G}_{2}$  &(1,0,0) &1& $\times$  &  $(\Z_2)^2$    & $(\Z_2)^2$\\
  \hline
 $\mathsf{G}_{3}$   &(2,0,0) &1& $\times$  &  $\Z_2$     & $(\Z_2)^3$ \\
  \hline
 {\color{black}$\mathsf{G}_{4}$} &(3,0,0) &1& {\color{black}$\times$}&  {\color{black}$\Z_2$}&{\color{black}$(\Z_2)^3$}\\
  \hline
  $\mathsf{G}_{5}$  &(0,1,0) &1& $\times$  &   $\Z_2$    &  $(\Z_2)^3$\\
  \hline
  $\mathsf{G}_{6}$ &(1,1,0) &1& $\times$  &  $(\Z_2)^2$    & $(\Z_2)^2$\\
  \hline
  $\mathsf{G}_{7}$   &(2,1,0) &1& $\times$  &       $\Z_2$    &  $(\Z_2)^3$ \\
  \hline
   $\mathsf{G}_{8}$  &(3,1,0) &1& $\times$  &      $(\Z_2)^2$    & $(\Z_2)^2$ \\
  \hline
   $\mathsf{G}_{9}$   &(0,0,1) &1& $\times$  &       $\Z_2$    &  $(\Z_2)^2$ \\
  \hline
  $\mathsf{G}_{10}$  &(1,0,1) &1& $\times$  &  $\Z_2$    & $\Z_2$ \\
  \hline
 $\mathsf{G}_{11}$   &(2,0,1) &1& $\times$  &   $\Z_2$   &  $(\Z_2)^2$ \\
  \hline
{\color{black}$\mathsf{G}_{12}$}   &(3,0,1) &1& {\color{black}$\times$}  & {\color{black} $\Z_2$}    & {\color{black} $\Z_2$}    \\
  \hline
 \hline
\end{tabular}
\caption{Classification of fSPT for Kramers-singlet fermions with $G_f=\Z_2^f\times G_b$ and $\mathcal{T}^2=\nu_2^b=1$. $G_b$ is generated by $\mathbf{g}$,  $\mathbf{h}$ and  $\mathcal{T}$, with  relations $\mathbf{g}^8=\mathbf{h}^2=\mathcal{T}^2=1$,  $\mathcal{T}\mathbf{g}\mathcal{T}^{-1}= \mathbf{g}^{1+2p}\mathbf{h}^q$ and $\mathcal{T}\mathbf{h}\mathcal{T}^{-1}= \mathbf{g}^{4t}\mathbf{h}$. 
}
\label{tab:singlet1}
\end{table}

\begin{table}[t]
\begin{tabular}{| c |c|c|c |c |c |}
\hline
\hline
 \,\,\,$G_f$\,\,\,  & ~$(p,q,t)$~ &~~$\mathcal{T}^2$~& Kitaev & complex fermion &\textit{  } bosonic \textit{  }  \\ 
  \hline
  $\mathsf{G}_{13}$   & (3,0,0) & $\mathbf{g}^4$ & $\Z_2$  &  $(\Z_2)^3$    & $\Z_2$\\
  \hline
  $\mathsf{G}_{14}$& (0,1,0) & $\mathbf{g}^2$ & $\Z_2$ & $(\Z_2)^2$ & $\Z_2$ \\
  \hline
   $\mathsf{G}_{15}$& (1,1,0) & $\mathbf{g}^4$ & $\Z_2$ & $(\Z_2)^3$ & $\Z_2$ \\
  \hline
   $\mathsf{G}_{16}$& (2,1,0) & $\mathbf{g}^2$ & $\Z_2$ & $(\Z_2)^3$ & $\Z_2$ \\
  \hline
  $\mathsf{G}_{17}$& (3,1,0)&$\mathbf{g}^4$ & $\Z_2$  &  $(\Z_2)^3$    & $\Z_2$\\
  \hline 
\hline
\end{tabular}
\caption{Classification of fSPT for Kramers-singlet fermions with $G_f=\Z_2^f\times G_b$ and $\mathcal{T}^2=\nu_2^b=\mathbf{g}^2$ or $\mathbf{g}^4$.  
In particular, $\mathsf{G}_{13}=\Z_2^f\times \Z_2\times  Q_{16}$.
Note that the KC decoration for $\Z_2^f\times Q_{16}$ is $\Z_2$, so the $\Z_2$ classification from KC decoration for $\mathsf{G}_{13}$ is purely from $Q_{16}$ and does not correspond to the one that leads to 3D Ising fTO. In addition, all other KC-decorated fSPTs do not correspond to the $K_f$ fSPT that leads to 3D Ising fTO.}
\label{tab:Q8}
\end{table}

We classify the 3D fSPT phases of all the groups listed in Table \ref{tab:singlet1} and \ref{tab:Q8}. 3D fSPT phases are constructed and classified in Refs.~\cite{QR18} using the decorated domain wall picture. The classification involves several layers of decoration, namely Kitaev-chain decoration, complex fermion decoration, and the intrinsic bosonic SPT phases. (There is an additional layer of $p+ip$ superconductor decoration. However, it is not relevant to our discussion.) The compatibility between decorations in different layers results in the so-called \textit{obstruction functions}. An fSPT state can be constructed if and only if the obstructions vanish. Technically, it is not easy to check whether the obstruction functions associated with a given decoration vanish. Hence, we adopt the computer codes implemented in GAP to do this job \cite{Ouyang_2021}.

We find that, for all groups in Table \ref{tab:singlet1}, no Kitaev-chain decoration is valid, i.e., Kitaev-chain decoration always leads to an obstructed fSPT construction. Hence, none of the fSPT phases found in Table \ref{tab:singlet1} reduces to the 3D Ising fTO after gauging the normal subgroup $\Z_2\times \Z_8$, because the latter results from Kitaev-chain decoration. Some of the cases are easy to understand. Take $\mathbf{r}=(0,0,0)$ for example, i.e., $\mathsf{G}_1$ in Table \ref{tab:singlet1}.  If an fSPT state of $\mathsf{G}_1$ symmetry is constructed based on Kitaev-chain decoration, it must also be an fSPT state of $K_f\subset \mathsf{G}_1$. As discussed above, the root state of 2D $\Z_2^f\times \Z_2$ fSPT, denoted as $[1]$, is decorated on the $\mathbf{g}$ DW. Under TRS, $\mathbf{g}$ DW remains as $\mathbf{g}$ DW, as determined by the group structure of $\mathsf{G}_1$. On the other hand, the root state $[1]$ is turned into the inverse state $[7]$ under TRS.  Therefore, the $K_f$ fSPT is not invariant under TRS action, meaning TRS with  $\mathbf{r}=(0,0,0)$ is not compatible with the $K_f$ fSPT. A similar simple argument exists for a few other cases, such as $\mathbf{r}=(1,0,0)$ and $(2,0,0)$. In fact, this argument is equivalent to the compatibility between three-loop braiding and TRS, discussed above.

For the groups in Table \ref{tab:Q8}, we do find valid (i.e. obstruction-free) Kitaev-chain decoration for constructing 3D fSPT phases. Nevertheless, these fSPT phases do not reduce to the 3D Ising fTO either. For $\mathbf{r}=(3,0,0)$ and $\nu_2^b=\mathbf{g}^4$, the group $G_f=\Z_2^f\times \Z_2\times Q_{16}$, where $Q_{16}$ is the generalized quaternion group with order 16,  with $\mathcal{T}^2=\mathbf{g}^4$, $\mathbf{g}^8=1$ and $\mathcal{T}\mathbf{g}=\mathbf{g}^7\mathcal{T}$.  We find that the fSPT phases with nontrivial Kitaev-chain decoration is protected by $Q_{16}$ alone. It is verified by turning off the $\Z_2$ symmetry and computing the classification of $\Z_2^f\times Q_{16}$ fSPT phases (see caption of Table \ref{tab:Q8}), which also contains a KC-decorated fSPT with $\Z_2^f\times Q_{16}$. Hence, by gauging $\Z_2\times \Z_8\subset \Z_2^f\times \Z_2\times Q_{16}$, we will not obtain the 3D Ising fTO. For $\mathbf{r}=(p,1,0)$  and $\nu_2^b = \mathbf{g}^{2}$ or $\mathbf{g}^4$, the group structures of $G_f$s are more complicated but all are order 32 groups.  We find that the only KC-decorated fSPT of each $G_f$ does not reduce to the KC-decorated fSPT of $K_f$ as well. This is seen by restricting the two-cocycle  $\tilde n_2$ of $G_f$ (accounting for KC decoration) to $\Z_2\times \Z_8$, which turns out to be associated with a trivial topological invariant $\Xi_{12}=0 $ mod 2. Meanwhile, the $n_2$ associated with the KC-decorated $K_f$ fSPT has a nontrivial topological invariant $\Xi_{12}=n_2(\mathbf{g},\mathbf{h})+n_2(\mathbf{h},\mathbf{g})$ =  1 mod 2. Combining the above results, we conclude that none of the fSPT phases we find become the 3D Ising fTO after gauging the normal subgroup $\Z_2\times\Z_8$.  That means TRS with $\mathcal{T}^2=1$ is incompatible with the 3D Ising fTO, consistent with the argument from the 3D SET viewpoint.

Finally, by employing the crystalline equivalence principle \cite{Else_gaugeSpatial}, we conclude that fermionic topological phase with $\mathcal{T}^2=1$ one-to-one corresponds to those with reflection symmetry $\mathcal{M}^2=\mathcal{P}_f$\cite{Witten_RMP,Else_gaugeSpatial}. Given that the Ising fTO enforces TRS with $\mathcal{T}^2=1$ to break, the reflection symmetry $\mathcal{M}^2=\mathcal{P}_f$ must break as well \footnote{When acting on Dirac fermions in 4d spacetime, the reflection symmetry acts as $\mathcal{M}^2=-1$ on single particle or $\mathcal{M}^2=P_f$ on many-body states. \cite{Zheyan23}}. Additionally, in 3D where parity symmetry acts as reflection and two-fold rotation symmetries, we conclude that parity symmetry with $P^2=1$ 
is enforced to break by such a 3D fTO as well.

\textit{Kramers-doublet fermions} ---
Here we show that the 3D Ising fTO  is compatible with  $\mathcal{T}^2=\mathcal{P}_f$. 
To do this, it is enough to find one obstruction-free fSPT example that contains $\Z_2\times \Z_8$ as a normal subgroup. The simplest example we find is associated with $\mathbf{r}=(3,0,0)$ and $\nu_2^a=\mathcal{P}_f$ and trivial $\nu_2^b$. More explicitly, the related symmetry is  $G_f= (\Z_4^{\mathcal{T}f}\rtimes \Z_8) \times \Z_2$ or equivalently $(\Z_{4}^{\mathcal{T}f}\times D_{16} \times \Z_2)/\Z_2^\mathcal{T}$ whose relations of the generators are given by $\mathbf{g}^8=\mathbf{h}^2=1$, $\mathcal{T}^2=\mathcal{P}_f$ and $\mathcal{T}\mathbf{g}\mathcal{T}^{-1}=\mathbf{g}^7,\mathcal{T}\mathbf{h}\mathcal{T}^{-1}=\mathbf{h}$. We summarize its classification in Table \ref{tab:spinhalf}. The KC decoration layer falls into $(\Z_2)^4$ classification, where one $\Z_2$ is protected solely by TRS, another by $D_{16}$, a third by jointly TRS and $\Z_2$, and the last one jointly by $\Z_2\times D_{16}$. The last $\Z_2$ reduces to the KC-decorated fSPT of $\Z_2 \times \Z_8$ when ignoring TRS. We have checked that it is indeed associated with the topological invariant $\Xi_{12}=1$ mod 2. Hence, it becomes the 3D Ising fTO if we gauge $\Z_2\times \Z_8$.  This phase can be realized through a real space construction by mapping to equivalent crystalline partners (see Supplementary Materials for more details).  

\begin{table}
\begin{tabular}{ c  |c |c |c }
\hline
\hline
 $G_f$ & Kitaev & complex fermion &bosonic \\ 
 \hline
 {\footnotesize $\Z_4^{Tf}\times \Z_2$} &  $(\Z_2)^2$  &  $ (\Z_2)^2 $ & $(\Z_2)^2 $\\  
\hline
 $\Z_4^{Tf}\rtimes  \Z_8$ &  $(\Z_2)^2$  &  $ (\Z_2)^3 $ & $(\Z_2)^2 $\\  
 \hline
 
 \hline
  $(\Z_4^{Tf}\rtimes \Z_8)  \times \Z_2$ & $(\Z_2)^4$  &  $(\Z_2)^6$    & $(\Z_2)^7$\\
  \hline
\hline
\end{tabular}
\caption{Classification of fSPT with Kramers-doublet fermions with $G_f= (\Z_4^{\mathcal{T}f}\rtimes \Z_8) \times \Z_2$. The first and second rows are classifications of fSPTs of its subgroups. All the three cases have another root fSPT from $p+ip$ decoration which is ignored since it is not relevant. }
\label{tab:spinhalf}
\end{table}

 \textit{Generality of 3D ESB} --- 
Finally, we discuss a general framework for studying ESB in 3D TO.  In TO, the global symmetry can have topologically distinct realizations, leading to diverse symmetry enrichment phenomena. These consist of the basis of SET phases.   By definition, ESB occurs if no anomaly-free 3D SET exists with a given 3D TO and a given global symmetry. 
The general theory of 3D ESB can be derived from 3D SET theory, which is not yet well-established, or it can alternatively be established by examining the related 3D SPT problem. 
Here, we discuss the latter method, whose primary assumption is that all 3D anomaly-free SETs, regardless of bosonic or fermionic ones, can be realized by partially gauging certain 3D SPTs  \cite{LanAB,LanEF}. 
For fTOs, we consider those that can be obtained by gauging $N$ of $N_f=\Z_2^f\times N$ fSPT.   
Denote the global symmetry of the SET and of the corresponding fSPT as $G_f$ and $\mathcal{G}_f$ respectively. The former contains $\Z_2^f$ in its center and  it is a central extension of $G_b$ (called bosonic symmetry) by $\Z_2^f$, described by $w_2\in \mathcal{H}^2(G_b,\Z_2^f)$.   The latter contains $N$ as a normal subgroup so that we can gauge it in $\mathcal{G}_f$ fSPT to obtain the fTO with the physical symmetry $G_f$ left.  These groups are subjected to the following short exact sequence:
\begin{align}
    1\rightarrow N_f\rightarrow \mathcal{G}_f \rightarrow G_b\rightarrow 1
\end{align}
defined by  a homomorphism $\rho$ from $G_b$ to $\text{Aut}(N_f)$ and a second cohomology class $\nu_2 \in \mathcal{H}_\rho^2(G_b, \mathcal{Z}(N_f))$ , which should satisfy the following physical conditions:
(i) the homomorphism $\rho$ only restricts to those in $\text{Aut}(N)$, so the cohomology class can split into $\nu_2=(\nu_2^a, \nu_2^b)$ with $\nu_2^a\in  \mathcal{H}^2(G_b, \Z_2^f)$ and $\nu_2^b\in \mathcal{H}^2_\rho(G_b,\mathcal{Z}(N))$ where $\mathcal{Z}(N)$ is the center of $N$; and (ii) the cohomology class $\nu_2^a  $ should be exactly the $w_2$ defined by $G_f$.
 If an anomaly-free 3D SET (with the given fTO and the given global symmetry $G_f$) exists, there is at least one 3D SPT protected by $\mathcal{G}_f$, which, when partially gauging $N$, yields this SET. Conversely, the absence of such a 3D $\mathcal{G}_f$ SPT indicates no anomaly-free 3D SET, which means ESB occurs.  So, to look for ESB of the given global symmetry $G_f$ by the given fTO, we check the classification of $\mathcal{G}_f$ fSPT with $\mathcal{G}_f$ defined by the above conditional group extensions, among which if there is no $\mathcal{G}_f$  fSPT such that gauging $N$ results in the given fTO, then there will be ESB. Finally, we note that, for bosonic SETs, the scheme is similar and simpler as there is no complication from $\Z_2^f$. 
 Further discussion on the general theory of 3D ESB will appear in Ref.\cite{SQ24b}.

\textit{Conclusion and discussion ---}
In summary, we demonstrate that $\mathcal{T}^2=1$ TRS is enforced to break by a kind of exotic 3D fTO, featuring Ising non-Abelian three-loop braiding statistics. However, the usual TRS with $\mathcal{T}^2=\mathcal{P}_f$ is still compatible with such a 3D fTO.
We also discussed a framework for exploring ESB for generic 3D fTOs. We further conjecture that parity violation in the Standard Model might arise from the vacuum being certain Ising 3D fTO, and it is a kind of ESB phenomenon.  

 \textit{Acknowledge} --- S.Q Ning thanks X.G Wen, H.T Lam and  A. Raz for enlightening discussions at the PITP summer school. This work is supported by funding from Hong Kong’s Research Grants Council (GRF No. 14308223, RFS2324-4S02, AoE/P-404/18, CRF C7012-21GF, CRS HKU701/24). S.Q.N. is supported by a CRF from the Research Grants
Council of the Hong Kong (No. C7037-22GF).  Y.Q. is supported by National Key R\&D Program of China (Grant No. 2022YFA1403402),  by National Natural Science Foundation of China (Grant No. 12174068) and by the Science and Technology Commission of Shanghai Municipality (Grant No. 23JC1400600). C.W. is supported by Research Grants Council of Hong Kong (GRF 17300220 and GRF 17311322) and National Natural Science Foundation of China (Grant No. 12222416).  

\bibliography{3DESB} 

\appendix
\newpage
\begin{widetext} 
\subsection{Group extension and constraints on $G_b$}
 \label{sec:app_automorphism}
We consider the $G_f$ is a group extension of $\Z_2^\mathcal{T}$ by $K_f=\Z_2^f\times \Z_2\times \Z_8$ given by a short exact sequence
\begin{align}
    1\rightarrow K_f \rightarrow G_f\rightarrow \Z_2^T\rightarrow 1.
\end{align}

In general, such a group extension is classified by two data: the group homomorphism $\rho: \Z_2^\mathcal{T}\rightarrow \text{Aut}(K_f)$ and $\mathcal{H}^2_\rho(\Z_2^\mathcal{T},K_f)$. 
 We do not consider the choice of $\rho$ such that $\rho_\mathcal{T}$ exchanges $P_f$ and elements in $\Z_2\times \Z_8$ because it will induce permutation between bosonic particles and fermionic particles.  We do not consider the TRS action will attach $\Z_2\times \Z_8$ elements with possible $P_f$ under which $\Z_2\times \Z_8$ is no longer a normal subgroup of $G_f$ as well.
So the group extension we are considering can be classified by (1) the group homomorphism $\rho: \Z_2^\mathcal{T}\rightarrow \text{Aut}( \Z_2\times \Z_8)$ and (2)  $(\nu_2^a,\nu_2^b)\in\mathcal{H}^2(\Z_2^\mathcal{T},\Z_2^f)\times \mathcal{H}^2_\rho(\Z_2^\mathcal{T},\Z_2\times \Z_8)$. 

Among these, the simplest data is $\nu_2^a$, classified by  $\mathcal{H}^2(\Z_2^\mathcal{T},\Z_2^f)=\Z_2$, reflecting  Kramers singlet or Kramers doublet fermions.
For the former, the fermionic symmetry $G_f$ is in form $\Z_2^f\times G_b$ with $G_b$ as a group extension of $\Z_2^\mathcal{T}$ by $\Z_2\times \Z_8$. {Below we only consider $\nu_2^a=1$, i.e., Kramers singlet fermions.}

Now we discuss  $\rho$, which values $\text{Aut}(\Z_2\times \Z_8) \cong \Z_2\times D_4$ \cite{autogroup}, whose elements are labeled by a vector $\mathbf{r}=(p,q,t)$ with TRS action $\rho_\mathcal{T}: (\mathbf{g},\mathbf{h})\rightarrow (\mathbf{g}^{1+2p}\mathbf{h}^q,\mathbf{g}^{4t}\mathbf{h})$ where $p=0,1,2,3$ and $q, t=0,1$. When both $q=t=1$ and all $p$, the corresponding automorphism is order-4 which can not be a TRS homomorphism. Therefore,  there are 11 order-2 subgroups and 1 trivial group that can become TRS homomorphisms, labeled by $p=0,1,2,3$ and $q\cdot t=0$.

To determine the final $G_b$, we still need  the two cocycle $\nu_2^b\in \mathcal{H}_\rho^2(\Z_2^\mathcal{T}, \Z_2\times \Z_8)$.    We find that there are as many as 28 different $G_b$s in total for all $\rho$ and $\nu_2^b$. On the one hand, we calculate all the 28 cases, which we found no case can reduce to $K_f$ fSPT for the related fTO. On the other hand, they can be  simply ruled out by the following consideration.  As the KC decoration is given by $\mathcal{H}^2(G_b,\Z_2)$, for some $\rho$ and $\nu_2^b$, there is no element in $\mathcal{H}^2(G_b,\Z_2)$ such that when restricting $G_b$ to $\Z_2\times \Z_8\subset G_b$ it can be reduced back to the $n_2 \in \mathcal{H}^2(\Z_2\times \Z_8,\Z_2)$, corresponding to  the  $K_f$ fSPT related to the 3D fTO. For those $G_b$ and related $G_f$ fSPT, we can not obtain the 3D fTO by partially gauging $\Z_2\times \Z_8\in G_f$. So we only need to focus on those $\rho$ and $\nu_2^b$ such that there exists at least one element in $\mathcal{H}^2(G_b,\Z_2)$ such that this reduction condition can be satisfied. So for simplicity in our calculations, we consider all elements in $\mathcal{H}^2(G_b,\Z_2)$, which may or may not reduce to $n_2$. Hence, we have those cases in Table \ref{tab:Q8}.

Now we explicitly check the constraint on the $\rho$  and $\nu_2^b$ from this reduction condition, which turns out to ruling out 11 cases, leaving 17 cases as presented in the main text.
For this purpose, it is convenient to use the formalism of Lyndon-Hochschild-Serre (LHS) spectral sequence developed in Ref.\cite{QR21}.  To be compatible with the language used there, from now on we rephrase our notation and  denote the group elements of $G_b$ as $a_{\mathcal{T}^\mathbf{n}}$ or simply $a_\mathbf{n}$ where the $\Z_2\times \Z_8$ elements are $a=(\mathbf{g})^{a_1}(\mathbf{h})^{a_2}$ or simply $a=(a_1,a_2)$,  and $\mathbf{n}=0,1$ denote the trivial or nontrivial element in $\mathcal{Z}_2^\mathcal{T}$.   From Ref.\cite{QR21},   a generic 2-cocycle of $G_b$ with $\Z_2$ coefficient can be parameterized as 
\begin{align}
    F^{a_\mathbf{b},b_{\mathbf{m}}}=F^{{}^{{\mathbf{nm}}}a,^{{\mathbf{m}}}b}+ F^{{}^{{\mathbf{nm}}}\nu_2^b(\mathbf{n},\mathbf{m}),{}^{{\mathbf{nm}}}a{}^{{\mathbf{m}}}b} +F^{{}^{{\mathbf{n}}}a,0_\mathbf{m}} +F^{0_\mathbf{n},0_\mathbf{m}} 
    \label{eq:2cocycle_para}
\end{align}
with the convention $F^{a,0_\mathbf{0}}=F^{0_\mathbf{0},0_\mathbf{m}}=F^{0_\mathbf{n},0_\mathbf{0}}=0$, $a,b\in \Z_2\times \Z_8$.  The notation ${}^\mathbf{n}a$ represents the permutation of $(\mathcal{T})^\mathbf{n}$ on element $a$, namely $(\rho_{\mathcal{T}})^\mathbf{n}(a)$ or simply $\rho_\mathbf{n}(a)$, and the ``0'' in notation $0_\mathbf{n}$  means the trivial element in $\Z_2\times \Z_8$.  More explicitly, we have  ${}^{\mathbf{1}}a=(a_1+q a_2, (2p+1)a_2+4ta_1)$,  which just follows from $(\mathbf{h})^{a_1} (\mathbf{g})^{a_2} \rightarrow (\mathbf{h}\mathbf{g}^{4t})^{a_1} (\mathbf{g}^{2p+1} \mathbf{h}^{q})^{a_2}=(\mathbf{h})^{a_1+qa_2}(\mathbf{g})^{(2p+1)a_2+4ta_1}$.
When either $\mathbf{n}$ or $\mathbf{m}$ takes the trivial element $\mathbf{0}$, then $\nu_2^b(\mathbf{n},\mathbf{m})$ is trivial in $\Z_2\times \Z_8$. 

 When restricting $G_b$ to $\Z_2\times \Z_8$, the value $ \mathbf{n}$ and $\mathbf{m}$ in $ F^{a_\mathbf{n},b_{\mathbf{m}}}$ are restricted to $\mathbf{0}$, i.e.,  $F^{a_{\mathbf{0}},b_{\mathbf{0}}}$ which follows the reduction condition should take the form as 
\begin{align}
    F^{a,b}=n_1^
    \alpha(a)n_1^\beta(b)=a_1 b_2 \, \text{ mod }\, 2
\end{align}

According to Ref.\cite{QR21}, in order to satisfy the 2-cocycle condition for $F^{a_\mathbf{n},b_{\mathbf{m}}}$, it is equivalent to satisfy the following conditions for $F^{a,0_\mathbf{n}}$ (denoted as $F_\mathbf{n}(a)$) and $F^{0_\mathbf{n},0_\mathbf{m}}$

\begin{align}
    F^{{}^{\mathbf{n}}a,{}^{\mathbf{n}}b}+F^{a,b} & =F_\mathbf{n}(b)+F_\mathbf{n}(a)+F_\mathbf{n}(ab)  \label{eq:constrain1}\\
    F^{{}^{{\mathbf{ml}}}a,{}^{{\mathbf{ml}}}\nu_2^b(\mathbf{m},\mathbf{l})}+ F^{{}^{{\mathbf{ml}}}\nu_2^b(\mathbf{m},\mathbf{l}),{}^{{\mathbf{ml}}}a}
 &=F_{\mathbf{ml}}(a)+F_{\mathbf{l}}({}^{\mathbf{m}}a)+F_\mathbf{m}(a)\label{eq:constrain2} \\
  F^{0_\mathbf{m}, 0_\mathbf{l}}+F^{0_\mathbf{n}, 0_\mathbf{ml}}+F^{0_\mathbf{n}, 0_\mathbf{m}}+F^{0_\mathbf{nm}, 0_\mathbf{l}}+F_{\mathbf{l}}({}^{{\mathbf{nm}}}\nu_2^b(\mathbf{n},\mathbf{m}))  
&= F^{{}^{{\mathbf{nml}}}\nu_2^b(\mathbf{n},\mathbf{ml}), {}^{{\mathbf{ml}}}\nu_2^b(\mathbf{m},\mathbf{l}) }
 + F^{{}^{{\mathbf{nml}}}\nu_2^b(\mathbf{nm},\mathbf{l}), {}^{{\mathbf{nml}}}\nu_2^b(\mathbf{n},\mathbf{m}) } 
    \label{eq:constraint3}
   \end{align}

\textbf{Cases with $q=0$} ---   Consider the automorphism with $q=0$ in $\mathbf{r}=(p,q,t)$. The left hand side of Eq. (\ref{eq:constrain1})  equals to zero modulo 2. Therefore, fixing $\mathbf{n}$, $F_\mathbf{n}(a)$ is a one cocycle in $\mathcal{H}^1(\Z_2\times \Z_8, \Z_2)$, which can be parameterized as
\begin{align}
    F_\mathbf{n}(a)=n_1(a)\mu(\mathbf{n})
\end{align}
with  $n_1\in \mathcal{H}^1(\Z_2\times \Z_8,\Z_2)$ and $\mu(\mathbf{0})=0$ due to $F^{a, 0_\mathbf{0}}=0$.  The  Eq.(\ref{eq:constrain2}) are simplified to  
\begin{align}
  &  a_1\nu_2^b(\mathbf{m},\mathbf{l})_2+a_2\nu_2^b(\mathbf{m},\mathbf{l})_1\nonumber \\
    & =n_1(a)\delta \mu (\mathbf{m},\mathbf{l})+[n_1(a)+n_1({}^{\mathbf{m}}a)]\mu(\mathbf{l})
\end{align}
  From this, we obtain $\nu_2^b(\mathbf{1},\mathbf{1})=(0,0)$ mod 2. (Remind that $\nu_2^b(\mathbf{1},\mathbf{1})$ here is just $\nu_2^b(\mathcal{T},\mathcal{T})$)   This can be used to evaluate the left side of Eq.(\ref{eq:constraint3}) which simply becomes zero, so  Eq.(\ref{eq:constraint3}) becomes the cocycle condition for $F^{0_\mathbf{n}, 0_\mathbf{m}}$ as 2-cocycle of $\Z_2^\mathcal{T}$, which does not have any requirement on $\nu_2^b$. 

To summarize, the $\nu_2^b$ is required to be in the form with
\begin{align}
    \nu_2^b(\mathcal{T},\mathcal{T})=(0, 2n_\mathcal{T}) \in \Z_2\times \Z_8
    \label{eq:w2_result1}
\end{align}
with $n_\mathcal{T}= 0,1,2,3$ mod 4.

Now we check the 2-cocycle condition, which, together with Eq.\ref{eq:w2_result1}, requires that $\delta \nu_2^b(\mathcal{T,T,T})=(0, 4p n_\mathcal{T})=(0,0)$ mod $(2,8)$. So if $p=0,2$, $n_\mathcal{T}$ can take $0,1,2,3$ while if $p=1,3$, $n_\mathcal{T}$ can take $0,2$. 

We still need to consider the  coboundary (gauge) transformation.  For the generic automorphisms, we can shift the two-cocycle by a coboundary $\nu_2^b(\mathcal{T},\mathcal{T})\rightarrow \nu_2^b(\mathcal{T},\mathcal{T}) \delta u(\mathcal{T},\mathcal{T})$ with \begin{align}
    \delta u(\mathcal{T},\mathcal{T})&={}^{\rho_\mathcal{T}}u(\mathcal{T})+u(\mathcal{T})-u({1})={}^{\rho_\mathcal{T}}u(\mathcal{T})+u(\mathcal{T}) \nonumber \\
    &=(0, (2p+2)u_2(\mathcal{T})+4tu_1(\mathcal{T})) \text{ mod } (2,8)
\end{align}
where 1-cochain $u=(u_1,u_2)\in C_\rho^1(\Z_2^\mathcal{T}, \Z_2\times \Z_8)$. 

So the first component of $\nu_2^b$ is unaffected. 

Through the coboundary transformation from $u_2(\mathcal{T})$, most cases can only take the trivial value 0 of $n_\mathcal{T}$  except for two cases $(3,0,0)$ and $(3,0,1)$. Further considering $u_1(\mathcal{T})$,  the value $n_\mathcal{T}=2$ for $(3,0,1)$ can be coboundary transformed into the trivial trivial $n_\mathcal{T}=0$, which means they represent two isomorphic group extensions.

\textbf{Cases with $q=1$} ---
We can assume $\nu_2^b(\mathcal{T},\mathcal{T})=(w_1,w_2) \in \Z_2\times \Z_8 $.  For this case,   the coboundary transformation for this $\rho_\mathcal{T}$  becomes 
\begin{align}
    &\delta u(\mathcal{T},\mathcal{T})=(u_2(\mathcal{T}) , (2p+2)u_2(\mathcal{T})) \text{ mod } (2,8)
\end{align}
So we can always choose $u_2(\mathcal{T})=w_1$ mod 2 so that $\nu_2^b(\mathcal{T},\mathcal{T})=(0,w_2)$. From now on, we assume that we have performed such a gauge transformation.  

The 2-cocycle condition leads $\nu_2^b(\mathcal{T},\mathcal{T})$ to be in the form of  $w_2=2n_\mathcal{T}$ mod 8 with a condition $4 p n_\mathcal{T}=0$ mod 8.  We can further perform a coboundary transformation with $u_2(\mathcal{T})=2u_\mathcal{T}$ without affecting  fixed $w_1=0$.   For $p=0$ or $2$,   $n_\mathcal{T}$ can take $0,1,2,3$, which, however, can further reduce to $0,1$ by further coboundary transformation with $u_\mathcal{T}=1$.  For $p=1$ or $3$,  $n_\mathcal{T}=0,2$ which is invariant under further coboundary transformation. Therefore, for $p=0$ or $2$, 
$\nu_2^b$ can take trivial and nontrivial, namely $\mathcal{T}^2=1$ or $\mathbf{g}^2$,  while for $p=1$ or 3,  $\nu_2^b$ can also take trivial and nontrivial,  namely $\mathcal{T}^2=1$ or $\mathbf{g}^4$.   

\subsection{Derivation of Eq.(\ref{eq:Theta})}
Now let us present the derivation of Eq.(\ref{eq:Theta}).
\begin{align}
  \Theta_{\rho(\Omega_1);\,\rho(\Omega_2)} 
   :=& 2\theta_{\rho(\Omega_1);\,\rho(\Omega_2)} \nonumber \\
     = & 2\theta_{\Omega_1\Omega_2^{4t};\, \Omega_2^{2p+1}\Omega_1^{q}} \nonumber \\
      = & 2\theta_{\Omega_1\Omega_2^{4t};\, \Omega_2^{2p+1}}+2\theta_{\Omega_1\Omega_2^{4t};\,\Omega_1^{q}}  \nonumber 
      \end{align}
      Note that the topological invariant is additive on the base loops. We can easily show that the second term  in the last line is equal to zero modulo $2\pi$ by noticing that (1) $qt=0$, (2) $\Omega_1$ is Abelian on the base loop $\Omega_1$  and (3)$\Theta_{\Omega_1;\Omega_1}=0$ mod $2\pi$.  Then we have
      \begin{align}
     \Theta_{\rho(\Omega_1);\,\rho(\Omega_2)}    
      = & 2\theta_{\Omega_1\Omega_2^{4t};\, \Omega_2^{2p+1}} \nonumber \\
   =&  (2p+1) \theta_{\Omega_1\Omega_2^{4t};\, \Omega_2^2}   \nonumber \\
     =&  (2p+1) (\theta_{\Omega_1; \Omega_2^2} +\theta_{\Omega_2^{4t}; \, \Omega_2^2}+\theta_{\Omega_1,\Omega_2^{4t}; \Omega_2^2})  \nonumber \\
     =&  (2p+1) (\theta_{\Omega_1;\, \Omega_2^2} + 16t^2\theta_{\Omega_2;\, \Omega_2^2}+ 4t\theta_{\Omega_1,\Omega_2;\, \Omega_2^2})  \nonumber \\
     =&  (2p+1) (\theta_{\Omega_1;\, \Omega_2^2} + 32t^2\theta_{\Omega_2;\, \Omega_2}+ 8t\theta_{\Omega_1,\Omega_2;\, \Omega_2})  \nonumber \\
     = &  (2p+1) (2\theta_{\Omega_1;\, \Omega_2}  ) + t (8\theta_{\Omega_1,\Omega_2;\, \Omega_2}) \nonumber \\
    =&  (2p+1) \Theta_{\Omega_1;\, \Omega_2}   + t \Theta_{\Omega_1,\Omega_2;\, \Omega_2}
 \end{align}
Some remarks are made for this derivation. Through the derivation,  we have used the fact that the topological invariant is additive on the base loops.  In the fourth and fifth lines, we have used the fact that on the base loop $\Omega_2^2$ both the loop $\Omega_1$ and $\Omega_2^{4t}$ are Abelian.  In the seventh line, we have used the fact that the topological invariant $\Theta_{\Omega_2;~\Omega_2}=8\theta_{\Omega_2;~\Omega_2}=0$. In the last line, we have used the definition that we have $\Theta_{\Omega_1,\Omega_2;\, \Omega_2}=8\theta_{\Omega_1,\Omega_2;\, \Omega_2}=\mu \pi$.  This topological invariant indicates the presence or not of a  Dijkgraaf-Witten type bosonic TO with $\Z_2\times \Z_8$ as gauge group, which is an Abelian TO.

Now we consider how  $ \Theta_{\Omega_2;\,\Omega_1} $ that can value $0$ or $\pi$  transforms when $t=0$
\begin{align}
     &\Theta_{\rho(\Omega_2);\,\rho(\Omega_1)} = 8  \theta_{\rho(\Omega_1);\,\rho(\Omega_2)}  \nonumber \\
     &=8\theta_{\Omega_2^{2p+1}\Omega_1^q;\Omega_1}\nonumber \\
&=8\theta_{\Omega_2^{2p+1};\Omega_1}+8\theta_{\Omega_1^q;\Omega_1}+8\theta_{\Omega_2^{2p+1},\Omega_1^q;\Omega_1} \nonumber \\
&=(2p+1)^2\Theta_{\Omega_2;\Omega_1}+4q \Theta_{\Omega_1;\Omega_1}+q(2p+1)\Theta_{\Omega_2,\Omega_1;\Omega_1} \nonumber \\
&=\Theta_{\Omega_2;\Omega_1}+q\Theta_{\Omega_2,\Omega_1;\Omega_1}
\end{align}
We note that on the base loop $\Omega_1$, the loop $\Omega_1$ is Abelian, namely it does not carry Majorana zero mode as does $\Omega_2$. Furthermore, we note the fact that $\Theta_{\Omega_2,\Omega_1;\Omega_1}=\pi$ mod $2\pi$. Therefore,  only $q=0$ is consistent.

\subsection{Review of $\Z_2^f\times \Z_2\times \Z_8$ fSPT phases and the corresponding three-loop statistics}

We review some properties of the 3D fSPT phases protected by $\Z_2^f\times \Z_2 \times \Z_8$ symmetry. It is known that 3D fSPT phases can be characterized by three-loop braiding statistics $\theta_{\alpha,\gamma}, \theta_{\alpha\beta,\gamma}, \dots$ after gauging the symmetry, which are defined between loops that are linked to a certain base loop $\gamma$ (see Refs.\cite{MC18_tlb,Zhou_Nonabelian}). In Abelian gauge theories, three-loop statistics can be well understood via dimensional reduction, by compactifying one of the three dimensions into a circle $S^1$ with a gauge flux $\phi_\gamma$ inserted in, where $\phi_\gamma$ is the gauge flux carried by the base loop $\gamma$. After that, one obtains a 2D fSPT phase, denoted as $\cal{S}(\gamma)$. (Strictly speaking, the 2D phase $\mathcal{S}$ only depends on the gauge flux $\phi_\gamma$. However, we denote it as $\mathcal{S}(\gamma)$ if no confusion is caused.) Then, the three-loop statistics $\theta_{\alpha,\gamma}, \theta_{\alpha\beta,\gamma},\dots $ correspond to the usual braiding statistics in the 2D theory $\mathcal{S}(\gamma)$. Equivalently, one may view the 2D fSPT state $\cal{S}(\gamma)$ as living on a membrane operator which creates the base loop $\gamma$. To describe $\Z_2^f\times \Z_2\times \Z_8$ 3D fSPT phases,  we need to consider $\mathcal{S}(\Omega_0)$, $\mathcal{S}(\Omega_1)$ and $\mathcal{S}(\Omega_2)$ for the three basic loops: the fermion parity loop $\Omega_0$, the $\Z_2$ unit flux loop $\Omega_1$, and the $\Z_8$ unit flux loop $\Omega_2$.

First of all, 2D fSPT phases with  $\Z_2^f\times \Z_2\times \Z_8$ are classified by $\Z_8\times \Z_{16}\times \Z_2\times \Z_2\times \Z_2$. We denote the root states by $A$, $B_1$, $B_2$, $C_1$ and $C_2$, respectively. Here, $A$ is protected solely by $\Z_2$ symmetry, $B_1$ and $B_2$ are protected solely by $\Z_8$ symmetry, and $C_1,C_2$ are protected jointly by $\Z_2$ and $\Z_8$ symmetries. We note that $A$ and $B_2$ are Majorana fSPT states, $B_1$ is a complex-fermion fSPT state (while the stacking state $B_1\otimes B_1$ is a bosonic SPT state), $C_1$ denotes the bosonic SPT state, and $C_2$ is a complex-fermion fSPT state. These 2D fSPT states can be characterized by the topological invariants $\Theta_{i},\Theta_{ij}, \Theta_{ijk}$ with $i,j,k=0,1,2$, defined in Ref. \cite{chenjie17_2Dinv}. We list the topological invariants here
\begin{align*}
    A: \quad & (\Theta_1,\Theta_{01}, \Theta_{001}) = \left(\pi/4, -\pi/2, \pi\right) \\ 
    B_1: \quad & (\Theta_2,\Theta_{02}, \Theta_{002}) = \left( \pi/8, \pi, 0\right) \\ 
    B_2: \quad & (\Theta_2,\Theta_{02}, \Theta_{002}) = \left(0,0, \pi\right) \\ 
    C_1: \quad & (\Theta_{12},\Theta_{012}) = \left(\pi, 0\right) \\ 
    C_2: \quad & (\Theta_{12},\Theta_{012}) = \left(0, \pi\right) 
\end{align*}

Next, we describe the theories $\mathcal{S}(\Omega_0)$, $\mathcal{S}(\Omega_1)$ and $\mathcal{S}(\Omega_2)$, which collectively characterize the 3D FSPT phase.  They should be certain stacks of the 2D root states. According to Ref. \cite{chenjie17_2Dinv}, we find that
\begin{align}
    \mathcal{S}(\Omega_0) = \left(C_2\right)^{ x}, \quad \mathcal{S}(\Omega_1)= \left(B_1\right)^{ 8y}\otimes  \left(B_2\right)^{ x} \otimes \left(C_1\right)^{ x} \otimes \left(C_2\right)^{ x}, \quad \mathcal{S}(\Omega_2) = A^x\otimes \left(C_1\right)^y \otimes \left(C_2\right)^{x} 
\end{align}
where $x=0,1,\dots,7$ and $y=0,1$, and $X^{z}$ means ``a stack of $z$ copies of root state $X$''. The values of $x$ and $y$ correspond to the 3D fSPT classification $\Z_8\times \Z_2$. Using the connection between three-loop statistics and 2D fSPT phases, we have that the 3D fSPT states are characterized by the following topological invariants:

\begin{table}[h!]
    \centering
    \begin{tabular}{c|c||c|c|c|c|c||c|c|c|c|c}
    \hline\hline
        $\Theta_{12,0}$ & $\Theta_{012,0}$ & $\Theta_{2,1}$ & $\Theta_{02,1}$ &  $\Theta_{002,1}$ & $\Theta_{12,1}$ & $\Theta_{012,1}$ & $\Theta_{1,2}$ & $\Theta_{01,2}$ &  $\Theta_{001,2}$ & $\Theta_{12,2}$ & $\Theta_{012,2}$   \\
        \hline   $0$ & $x \pi$ & $y\pi$ & 0 & $x\pi$ & $x\pi$ & $x\pi$ &  $x\pi/4$ & $ -x\pi/2$ & $x\pi$ & $y\pi$ & $x\pi$ \\
    \hline\hline
    \end{tabular}
\end{table}

\noindent Other topological invariants are completely determined by the ones above. We are interested in 3D topological orders obtained by gauging Majorana fSPTs, so we only consider $x=1,3,5,7$ and $y=0,1$.  In this cases, we have
\begin{align}
    \mathcal{S}(\Omega_0) = C_2, \quad \mathcal{S}(\Omega_1)= \left(B_1\right)^{ 8y}\otimes  B_2 \otimes C_1 \otimes C_2, \quad \mathcal{S}(\Omega_2) = A^x\otimes \left(C_1\right)^y \otimes C_2
\end{align}
where we have used the fact that $B_2, C_1, C_2$ are of order two under stacking.

\section{Topological crystalline  phases corresponding to the compatible case}

\begin{table}[h]
\begin{tabular}{  c | c |c |c }
\hline
\hline
 $G_f$ & Kitaev & complex fermion &bosonic \\ 
 \hline
 {\footnotesize $\Z_4^{Tf}\times \Z_2$}  &  $(\Z_2)^2$  &  $ \Z_2 $ & $(\Z_2)^2 $\\  
\hline
\hline
\end{tabular}
\caption{Classification of 2D $\Z_4^{\mathcal{T}f} \rtimes \Z_8$ fSPT.}
\label{tab:2D}
\end{table}

Here we discuss how to construct the topological crystalline phases that corresponding to the $\Z_4^{\mathcal{T}f} \rtimes \Z_8\times \Z_2 $ fSPT that can reduce to the $\Z_2^f\times \Z_8\times \Z_2$ non-Abelian fSPT. We choose to transform to $\Z_2$ subsymmetry into the $C_2$ rotation symmetry, so that the topological crystalline phases are protected by $\Z_4^{\mathcal{T}f} \rtimes \Z_8\times C^{-}_2 $ symmetry, where $C_2^-$ here we mean the square of the $C_2$ rotation equals to fermionic parity. To see the construction, we first make two remarks. The first one is that the cell decomposition for $C_2$ rotation system in 3D consists of a 1D spatial line as the rotation center and two 2D planes that intersect at the rotation center and can be transformed into each other by rotation, and two 3D cells that are also transformed into each other by rotation.  To construct $\Z_4^{\mathcal{T}f} \rtimes \Z_8\times C^{-}_2 $ symmetric topological phases, we can place 1D, 2D, or 3D topological phases with certain symmetry on the 1D, 2D, or 3D cell accordingly. The second remark is that the classification of 2D $\Z_4^{\mathcal{T}f} \rtimes \Z_8$ fSPT is summarized in Table.\ref{tab:2D} where  we can see the classification from KC decoration is $(\Z_2)^2$: one $\Z_2$ is purely protected by TRS with $\mathcal{T}^2=\mathcal{P}_f$ while the other is purely by $\Z_8$. Now we claim that the decoration of 2D $\Z_4^{\mathcal{T}f} \rtimes \Z_8$ fSPT characterized by the purely $\Z_8$ KC decoration   on the 2D cell gives the topological crystalline phases we are looking for. As such a 2D state is $\Z_2$ classified, so along the rotation line there are two copies of 1D edge theory of such a 2D fSPT, which together is 't Hooft anomaly-free, so that they can be symmetrically gapped out to complete the crystalline phase construction.  

We also note that another compatible example exists with the TRS action $\mathbf{r}=(1,0,1)$. We further note that two scenarios with automorphisms $\mathbf{r}=(3,0,0)$ and $(3,0,1)$ are consistent with this non-Abelian TO and are both isomorphic to $G_f=(\Z_4^{Tf}\times \Z_2\times D_{16})/\Z_2^T$.

\end{widetext}

\end{document}